\documentclass[jeosman,preprint,fleqn,showpacs,showkeys]{revtex4}
\usepackage{amssymb,amsmath,bm}
\usepackage[pdftex]{graphicx}

\begin{document}

\title{Conditional Generation Scheme for a High-fidelity Yurke-Stoler States
by Mixing Two Coherent Beams with Squeezed Vacuum State}

\author{Sun-Hyun Youn\footnote{email: sunyoun@chonnam.ac.kr, fax: +82-62-530-3369}}
\address{Department of Physics, Chonnam National University, Gwangju 500-757, Korea}

\begin{abstract}

  The numerical conditions to generate a high-fidelity  Yurke-Stoler
states ($|\alpha>+  e^{i \psi} |-\alpha>$ )were found for two
cascade-placed beam splitters with one squeezed state input and two
coherent state inputs. Controlling the amplitude and the phases of
beams, allows for various Yurke-Stoler states to be manipulated with
ultra high-fidelity, and the expected theoretical fidelity is of
more than $0.9999$.



\end{abstract}


\maketitle

\section{Introduction}

  The generation of specific quantum states  at a high-fidelity  plays an
essential role in quantum information science\cite{Mskim08,Fabio06}.
Manipulating nonclassical light that is originally generated by a
nonlinear interaction between the light fields and  the medium has
been an effective   method for engineering the desired non-classical
photonic state in a quantum process\cite{wang, Fabio06}.

 The manipulation of the nonclassical state usually adopts a
conditional state-preparation scheme that takes advantage of a
strong nonlinearity induced by the quantum measurement,  even at the
level of a single photon. By measuring one of the entangled photons
produced in the parametric down conversion, the arbitrary
superposition of vacuum  and single-photon states has been
accomplished\cite{steinberg02}  and a variant combination of
coherent states and Fock states are conditionally  created and
analyzed \cite{alex01,Grangier06,alex02,Zavatta04,alex10}.

Generating nonlinear photonic-states, such as Schrodinger states,
using a squeezed light source  by means of conditional measurements
on a beam splitter has also  been   extensively studied, both
theoretically and experimentally\cite{dakna97, cerf05,ourjou06,
Nielsen, Wkui}.

\begin{figure}[htbp]
\centering
\includegraphics[width=5cm]{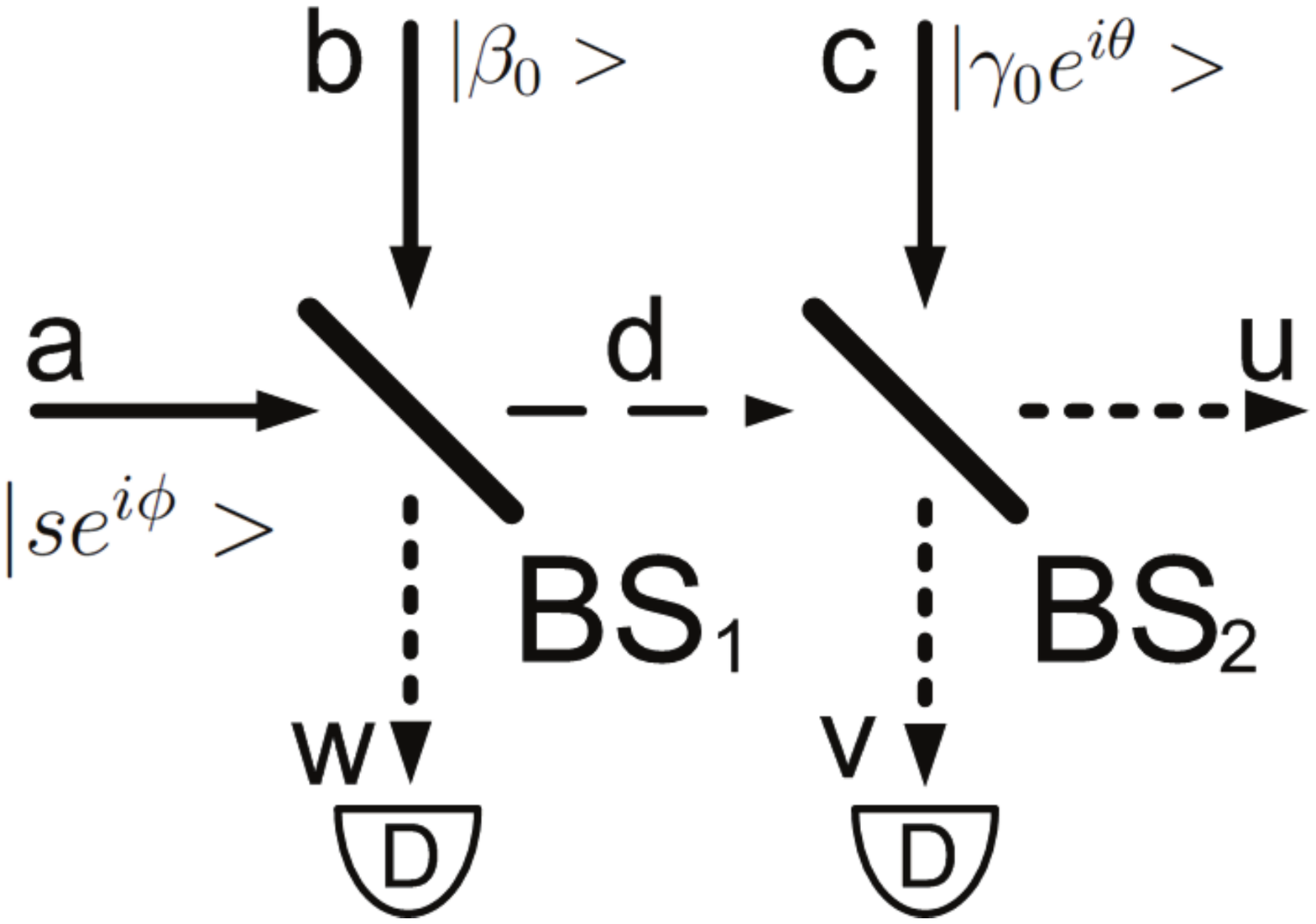}
\caption{Schematic diagram of the Yurke-Stoler state generation. One
squeezed state is in the input mode $a$ ($ |s e^{i \phi}> $ ), and
two coherent states ($ |\beta_0 = 0 >, |\gamma_0 e^{i \theta}> $)
are in the input modes $b$, and $c$.  BS: Beam Splitter. D:
Detector.} \label{expsetupA}
\end{figure}

In our proposed system, the junction of the field coherent state is
added to the non classical state \cite{alexprl02}, and the photon is
subtracted from the squeezed vacuum\cite{Gerrits}. We added two
coherent beams with cascade placed beam splitters,  as seen in Fig.
\ref{expsetupA}.   The two beamsplitters and two coherent beams give
us a degree of
   freedom to control the output in a highly nonclassical manner.
We characterize  the single output from the three input
    beams with a simultaneous detection of
two photo detectors.  Our system has a great advantage in that it
can generate a high-fidelity Fock-state \cite{youn13, youn14}.

In this article, we study the condition to generate  high-fidelity
Yurke-Stoler state ($|\alpha>+  e^{i \psi} |-\alpha>$ )
\cite{Yurke86}. The well-known Schrodinger-cat states are a special
case of the Yurke-Stoler state, and recently, the Schrodinger-cat
state has been generated from classical radiation in  the microwave
region through a subtraction measurement \cite{Govia}, and a
macroscopic Schrodinger-cat state has been  generated in a
nanomechanical resonator \cite{Zhang}. Basically two targets have
been set to generate a  Schrodinger-cat state. The first one is to
obtain a giant Schrodinger cat state (high $ \alpha $), and the
second  is to generate a high-fidelity one.

  The present paper is organized as follows. In Section II, we
 introduce our system with two cascade placed beam
 splitters, with one squeezed state and two coherent state
 inputs.   In Section III, we explicitly calculate the probability of the
  amplitude when two detectors at the output port
 simultaneously  detects a single photon, and then we find the condition for which
 the output port generates a  high-fidelity Yurke-Stoler state in
 Section IV.  To obtain an optimal state,  we tested the square of the  Wigner function differences,
  which are more sensitive  than the fidelity.  In Section V, we
 summarize the  main results and discuss the experimental implementation.

\section{Two cascade placed beam splitters}
Let a squeezed vacuum state $|\xi> $ be in the input mode $a$, and
the two coherent states, $|\beta>$ and $|\gamma>$,  be in the input
modes $b$ and $c$, as seen in the experimental set up in Fig.
\ref{expsetupA}.  Then, the input state $|\xi> , |\beta>, |\gamma>$
can be expressed in the number-state representation \cite{Loudon}:
\begin{eqnarray}
 |\xi,\beta,\gamma>
  = e^{-\frac{1}{2} (|\beta|^2+|\gamma|^2 )} \sum_{n_{\xi} = 0, n_{\beta} =0, n_{\gamma} =0
  } C_{n_\xi}
 \frac{\beta^{n_{\beta}}} { \sqrt{n_{\beta}!}}
 \frac{\gamma^{n_{\gamma}}} { \sqrt{n_{\gamma}!}}|n_{\xi}>_{a}
    |n_{\beta}>_{b}|n_{\gamma}>_{c}, \label{coherentSt}
\end{eqnarray}
where $C_{n_{\xi}}$ is the coefficient of the squeezed vacuum with
squeezing parameter,  $s e^{i \phi} $, and is zero for all odd
values of $n_{\xi}$  and nonzero only for an even value of
$n_{\xi}$. The nonzero values of  $C_{n_{\xi}}$ for even values of
$n_{\xi}$ become \cite{Loudon}: 
\begin{eqnarray}
  C_{n_\xi } =  \frac {\sqrt{n_\xi !}}{\sqrt{\cosh s} \frac{n_\xi }{2}!}
  (- \frac {1}{2} e^{i \phi} \tanh s )^{\frac{n_\xi }{2}}.   \label{sV}
\end{eqnarray}

With the experimental set up in Fig. \ref{expsetupA}, the three
creation operators $ {\hat a} ^{\dagger}, {\hat b} ^{\dagger}, {\hat
c} ^{\dagger}$ are written in terms of three creation operators $
{\hat w} ^{\dagger}, {\hat v} ^{\dagger}, {\hat u} ^{\dagger}$.
Using the operator relation \cite{Teich1989}, we can obtain the
relations between the input modes and the output modes as the
following,
\begin{eqnarray}
 \left(\begin{array}{c}
{\hat a}^\dagger  \\
{\hat b}^\dagger  \\
\end{array}\right)
 = \left(\begin{array}{cc}
 t_1  e^{ -i \phi_{\tau_1} } &  \sqrt{1- {t_1}^2} e^{ -i \phi_{\rho_1} } \\
-\sqrt{1- {t_1} ^2} e^{i \phi_{\rho_1}  } & t_1  e^{i \phi_{\tau_1} } \\
\end{array}\right) \left(\begin{array}{c}
{\hat d}^\dagger  \\
{\hat w}^\dagger  \\ \end{array}\right)  \label{BSab}
\end{eqnarray}
\begin{eqnarray}
 \left(\begin{array}{c}
{\hat d}^\dagger  \\
{\hat c}^\dagger  \\
\end{array}\right)
 = \left(\begin{array}{cc}
 t_2  e^{ -i \phi_{\tau_2} } &  \sqrt{1- {t_2}^2} e^{ -i \phi_{\rho_2} } \\
-\sqrt{1- {t_2} ^2} e^{i \phi_{\rho_2}  } & t_2  e^{i \phi_{\tau_2} } \\
\end{array}\right) \left(\begin{array}{c}
{\hat u}^\dagger  \\
{\hat v}^\dagger  \\ \end{array}\right).  \label{BScd}
\end{eqnarray}
Then the input states in Eq. \ref{coherentSt} can be written as
number-state representations of the output modes ($u, v, w$) as
follows \cite{youn13}:
\begin{eqnarray}
 |\xi,\beta,\gamma>
  &=& e^{-\frac{1}{2} (|\beta|^2+|\gamma|^2 )} \sum_{n = 0, l =0,m =0
  } C_{n} \frac{({\hat a}^{\dagger})^n } {\sqrt{n!}} \frac{(\beta {\hat b}^{\dagger})^l } {l!}
  \frac{(\gamma {\hat c}^{\dagger})^m } {m!}  |0>_{a}|0>_{b}|0>_{c}\\
  &=& e^{-\frac{1}{2} (|\beta|^2+|\gamma|^2 )} \sum_{n = 0, l =0,m =0
  } C_{n} \frac{(q_{a}^{u}  {\hat u}^{\dagger}+q_{a}^{v}  {\hat
v}^{\dagger}+q_{a}^{w}  {\hat w}^{\dagger})^{n}}{\sqrt{n!}} \nonumber \\
 &\times &  \frac{(\beta( q_{b}^{u}  {\hat u}^{\dagger}+q_{b}^{v}  {\hat
v}^{\dagger}+q_{b}^{w}  {\hat w}^{\dagger} ) )^l} { l !}
 \frac{(\gamma (q_{c}^{u}  {\hat u}^{\dagger}+q_{c}^{v}  {\hat v}^{\dagger}))^{m}} {m!}|0>_{u}
    |0>_{v}|0>_{w
    }. \label{StOp2}
\end{eqnarray}
Where $q_{\mu}^{\nu}$ ($\mu= a,b,c, \nu=u,v,w$) represents the
relations between the operators in the input modes (${\hat
a}^\dagger ,{\hat b}^\dagger, {\hat c}^{\dagger}$) and those in the
output modes (${\hat u}^\dagger ,{\hat v}^\dagger, {\hat
w}^{\dagger}$) as follows:
\begin{eqnarray}
\{ q_{a}^{u} , q_{a}^{v} , q_{a}^{w}  \} &=& \{  e^{- i (
\phi_{\tau_1}+\phi_{\tau_2} )} t_1
 t_2 ,  e^{- i ( \phi_{\rho_2}+ \phi_{\tau_1})} t_1 \sqrt{1-{t_2}
^2},  e^{-i  \phi_{\rho_1}} \sqrt{1-{t_1} ^2}  \} \nonumber \\
\{ q_{b}^{u}, q_{b}^{v}, q_{b}^{w} \} &=& \{ e^{ i \phi_{\rho_1}}
\sqrt{1-{t_1} ^2}  e^{ -i \phi_{\tau_2}} t_2, - e^{ i \phi_{\rho_1}}
\sqrt{1-{t_1} ^2}  e^{-i \phi_{\rho_2}} \sqrt{1-{t_2} ^2} , e^{i
\phi_{\tau_1}} t_1 \}\nonumber \\
\{  q_{c}^{u} ,q_{c}^{v} \} &=&  \{  -e^{ i \phi_{\rho_2}}
\sqrt{1-{t_2} ^2},  e^{i  \phi_{\tau_2}} t_2 \} \label{eqOPabc}.
\end{eqnarray}
Using the trinomial coefficients, $|\xi,\beta,\gamma> $ in Eq.
\ref{StOp2}  becomes
\begin{eqnarray}
 |\xi,\beta,\gamma>
  &=& e^{-\frac{1}{2} (|\beta|^2+|\gamma|^2 )} \sum_{n = 0, l =0,m =0} C_{n} \beta^{l}\gamma^{m} \frac{\sqrt{n!}}{n_u ! n_v ! n_w !}
   \frac{1}{l_u ! l_v ! l_w !}
\frac{1}{m_u ! m_v! }  \nonumber \\
   & \times &
(q_{a}^{u})^{n_u}  (q_{a}^{v})^{n_v} (q_{a}^{w})^{n_w}
(q_{b}^{u})^{l_u}  (q_{b}^{v})^{l_v}
(q_{b}^{w})^{l_w} (q_{c}^{u})^{m_u}  (q_{c}^{v})^{m_v} \nonumber \\
&\times &
 ({\hat u}^{\dagger})^{n_u + l_u + m_u}
 ({\hat v}^{\dagger})^{n_v + l_v + m_v}
 ({\hat w}^{\dagger})^{n_w + l_w }
 |0>_{u} |0>_{v}|0>_{w}, \label{StOp2}
\end{eqnarray}
where the $n'$ summation indicates all summations for  non negative
numbers $n_u$, $n_v$, and $n_w$  such that  $n_u + n_v + n_w = n $.
After collecting all terms that satisfy $N_u = n_u + l_u + m_u $,
  $N_v = n_v + l_v + m_v $, and $N_w = n_w + l_w$ over all $n, l$,
  and $m$, Eq. \ref{StOp2} can be written with new coefficients $C(N_u, N_v, N_w )$
  as follows
\begin{eqnarray}
 |\xi,\beta,\gamma>_{uvw}
  =  \sum_{N_u = 0, N_v = 0, N_w = 0 } C(N_u , N_v ,  N_w)
   |N_u > |N_v > |N_w>.  \label{simpleEq}
\end{eqnarray}
If two detectors in the $v$ and $w$ modes simultaneously detect a
single photon, the  probability of finding $n$ photons in the output
mode $u$ becomes
\begin{eqnarray}
|C_{s} (n)|^2  \equiv |C( N_u =n, N_v =1, N_w = 1 )|^2 .
\label{eqcs}
\end{eqnarray}

\section{Wigner function optimization}

The probability of finding $n$ photons in the output mode $u$ when
two detectors in the  $v$ and $w$ modes simultaneously detect a
single photon, $C_{s}$ in Eq. \ref{eqcs},  can be explicitly written
as a function of the amplitudes of the three input beams , the phase
differences of three beams, and the transmittances of the two beam
splitters \cite{youn13,youn14}. In order to generate Yurke-Stoler
states\cite{Yurke86}, we calculate the relative coefficients $C_{s}
(n) /C_{s} (0)$  as follows
\begin{eqnarray}
\{ \frac{C_{s}(1)}{C_{s}(0)} ,\;\;
\frac{C_{s}(2)}{C_{s}(0)},\;\;...\}
 = \{   e^{i \theta } \gamma_0 \frac{2 t_2 ^2 -1}{\sqrt{1- t_2 ^2}},\;\; \frac{1}{\sqrt{2}}
 \{e^{2 i \theta}(1-3 t_2 ^2 ) \gamma_0 ^2 - 3 e^{ i \phi} t_1 ^2 t_2 ^2 \tanh{s}  \},\;\;...  \}
 \label{EXPterms}
\end{eqnarray}
To find a simple solution, we set the amplitude of the $|\beta_0>$
to zero. Although, it's not possible to match all the coefficients,
we can optimize the conditions for the generation through a
numerical optimization. At first, we attempted to increase the
fidelity of the tested state. The fidelity ($F$) between the two
pure state $ |s_1
>$ and $|s_2
> $.  is defined by an overlap $|< s_1 | s_2
>| $ such as,
\begin{eqnarray}
F  &=& |\sum_{n,m} a^*_n  b_m   <n|m>| \nonumber \\
&=& | \sum _{n} a^* _n b_n |,
 \label{fidelity}
\end{eqnarray}
where we used $|s_1 > = \sum_n a_n |n> $ and $|s_2 > = \sum_m b_m
|m> $.

On the other hand, the Yurke-Stoler state is very sensitive to the
relative phase of each Fock state. So we calculated the Wigner
function of each trial state and then minimized the absolute square
of the difference between two Wigner functions of the Yurke-Stoler
state and target states. This method is more efficient than the
method using the fidelity in our numerical computer algorithm.

The Wigner function of the Yurke-Stoler state $W_{YS}$  is
\begin{eqnarray}
W_{YS} (x,p,\alpha,\zeta ) = \frac{1} {\pi}  \int e^{ 2i p y}
\psi_{YS} (x-y,\alpha , \zeta ) \psi_{YS}^* (x+y,\alpha ,\zeta) dy,
\label{WYS}
\end{eqnarray}
where the Yurke-Stoler state $\psi_{YS}$ is
\begin{eqnarray}
\psi_{YS}(x_,\alpha,\zeta ) =  <x|(|\alpha> + e^{i \zeta} |-\alpha>
). \label{YS}
\end{eqnarray}
For the trial function $g(x)$
\begin{eqnarray}
g(x) &=& \sum_{n=0} C_s (n) \psi_n (x),
\end{eqnarray}
with  $\psi_n (x) = <x|n>  $, the Wigenr function of the trial
functions $W_{g}$ becomes
\begin{eqnarray}
W_g (x,p) &=&  \frac{1} {\pi}  \int e^{ 2i p y} g(x-y) g^*(x+y) dy.
\label{wignerG}
\end{eqnarray}

Then the absolute square of the difference between the two Wigner
functions $D_w $ is
\begin{eqnarray}
D_w &=& 2 \pi  \int_{x,p} |W_{YS}(x,p,\alpha,\zeta) - W_g (x,p)|^2
dp dx \nonumber \\
&=&  1- 2 \pi   \int_{x,p} W_{YS} (x,p,\alpha,\zeta) W_g (x,p)  dp
dx. \label{errorInt}
\end{eqnarray}
The second term can be expanded as follows,
\begin{eqnarray}
& & 2 \pi  \int_{x,p} W_{YS}(x,p,\alpha,\zeta)  \{ \frac{1} {\pi}
\int_{y}
e^{ 2i p y}  g(x-y) g^*(x+y) dy \}  dp dx \nonumber \\
&=&  2 \pi   \int_{x,p} W_{YS} (x,p,\alpha,\zeta)  \{ \frac{1} {\pi}
\int_{y} e^{ 2i p y}  \sum_{n=0} C_s (n) \psi_n (x-y)
 \sum_{m=0} C_s^* (m) \psi_n^* (x+y) dy \} dp dx \nonumber \\
&=& 2  \sum_{n=0,m=0}  C_s (n) C_s^* (m) I (n,m,\alpha,\zeta).
  \label{errorInt5}
\end{eqnarray}
Algebraic manipulation leads to
\begin{eqnarray}
I(n,m,\alpha,\zeta ) &=&  \int_{x,p,y} e^{ 2i p y}
W(x,p,\alpha,\zeta)   \psi_n (x-y)
  \psi_n^* (x+y) dy  dp dx  \label{intTable} \\
  &=& \frac{ \{(-1)^n + e^{i \zeta}\}
  \{1+(-1)^m e^{ i \zeta}\} \alpha^{n+m}}
  { 2(1+2 e^{2 \alpha^2 + i \zeta} + e^{2 i \zeta}) \pi
  \sqrt{n!}\sqrt{m!}} e^{\alpha^2}
\label{intTableForm}
\end{eqnarray}

Finally, the closed integral formula in Eq. \ref{intTableForm} can
be used to efficiently minimize the  absolute square of the
difference between the two Wigner functions $D_w $.

\section{High purity cat-sate generation.}

 The generation of various nonclassical photonic-states through the use of a
squeezed light source and conditional measurements on a beam
splitter has  been  extensively studied, both theoretically and
experimentally.  We use squeezed vacuum states, combine the addition
and  subtraction of photons\cite{youn13}.  Using the cascaded beam
splitter as  used by Bimbard et al. \cite{alex10} ( Fig.
\ref{expsetupA}),  we can control the probabilities of the generated
quantum states.  The numerical conditions necessary to generate
high-purity $|1>$, $|2>$, and $|1>+r e^{i \psi} |2>$ states were
found with a value of  more than $1000$ for the expected theoretical
signal-to-noise ratio \cite{youn14}.  The two beam splitters and two
coherent beams with two detectors in the output port give us a
degree of freedom to control the output, which is highly
nonclassical.

We present the conditions to generate the Yurke-Stoler state with
$\alpha = 1$ and $\zeta = 0, \frac{2}{3} \pi, \frac{4}{3} \pi $, in
Table \ref{YS1}. The interesting thing is that if we only change the
relative phase of the input beams ($\phi $, and $\theta$), we can
generate three states
\begin{eqnarray}
 \{|\alpha_1 >+  | -\alpha_{1}>,\;\;|\alpha_1 >+ e^{i \frac{2}{3} \pi}
|- \alpha_{1}>,\;\; |\alpha_1 >+ e^{i \frac{4}{3} \pi} |-\alpha_{1}>
\},
 \label{three states}
\end{eqnarray}
where $ \alpha_{1} = 1 $.  The Yurke-Stoler state $|\alpha_1 >+
|-\alpha_{1}> $ can also be generated by blocking the $|\gamma_0
e^{i \zeta}>$ state and setting the phase differences. The phase
differences seem meaningless if the two coherent beams are removed,
but the actual phase differences contains all the phases of beam
splitters ($ \phi_{\rho_1}, \phi_{\rho_2},\phi_{\tau_1}, \phi_{
\tau_2} $) \cite{youn13}. Therefore the phase differences should be
controlled and checked through the very weak coherent light inputs
in the actual experiments.
 The fidelity of the three  Yurke-Stoler states are greater than $0.995$, and the
absolute difference of the Wigner states is of less than $0.01$.

With fixed amplitudes for the input states and the transmittances of
the two beam splitters, we obtain similar quantum states according
to the relative phases of the two beams. The last column in Table
\ref{YS1} , $ Log(P) $, shows the logarithmic value of the
 generating probability. The meaning of the probability is that,
  for a condition where the amplitudes
 of the input beams and the fixed transmittance value are given, there is
 a chance to simultaneously
 detect a pair of single photons  at each detector. If we consider the
 train of the input pulses, there is approximately 1 chance to generate
a state (
  $|\alpha_{1} >+  e^{i 2 \frac{\pi}{3}} | -\alpha_{1}
> $)  with a fidelity of  $0.997$ for every 1950 pulses.

\begin{table}
\centering \caption{($|\alpha_{1} >+  e^{i \zeta} | -\alpha_{1}
> $)-state generating conditions.} \label{YS1}
\begin{tabular}{c||c|c|c|c|c|c|c|c|r}
\hline $\zeta$  & $s$  & $\gamma_0$  & $t_1$ & $t_2$  & $\phi$ &
$\theta $ &  $F$   & $ Log (D_w)$ & $Log(P)$\\ \hline \hline
  $ 0 $  & $1 \over 2 $  & 0.0   & 0.797  & 0.997 &  $\pi$   & 0 & 0.995 &-2.02 & -3.29 \\ \hline
   $ {2 \over 3} \pi $  &$1 \over 2 $ & 0.137   & 0.797  & 0.997 & $\pi$     &- $  \frac {1}{2} \pi $ & 0.997 &-2.29 & -3.29\\ \hline
     $ {4  \over 3} \pi $  &$1 \over 2$  & 0.137   & 0.797  & 0.997 & -$\pi$    & $ \frac{1}{2}  \pi  $  & 0.997 &-2.23 & -3.29\\ \hline
     \hline
\end{tabular}
\end{table}
\begin{figure}[htbp]
\centering
\includegraphics[width=15cm]{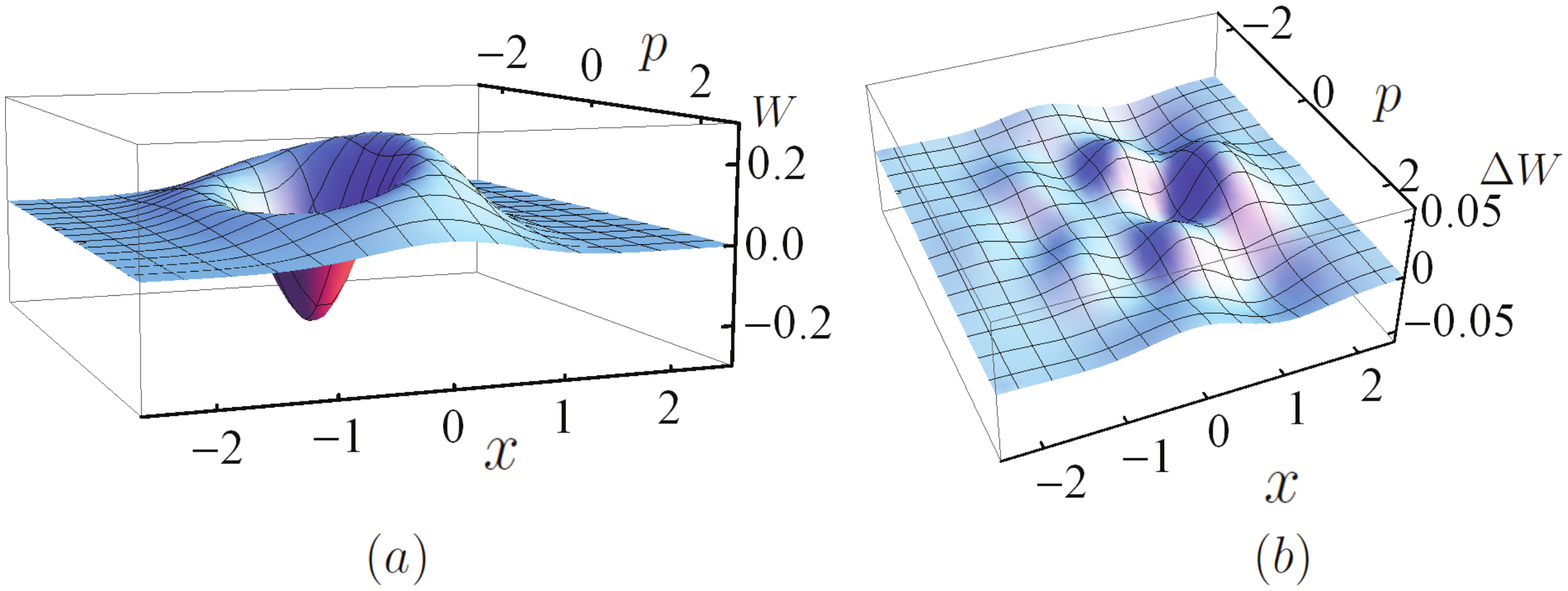}
\caption{(a) Wigner function of the generated states ( $\zeta =
\frac{4}{3} \pi $ ) under the setup in Table \ref{YS1}. (b)
Difference of the Wigner function between the state in (a) and the
Yurks-Stoler state ($|\alpha_{1} >+  e^{i \frac{4}{3} \pi}
|-\alpha_{1}> $) .} \label{picABagain}
\end{figure}

In Fig. \ref{picABagain} (a), we plotted the Wigner function of the
generated state by using the numerical parameters in Table
\ref{YS1}. The Wigner functions is calculated for the state with
$\zeta = \frac{ 4}{3} \pi$. In Fig. \ref{picABagain} (b), we plotted
the difference between the two Wigner functions for the state in
Fig. \ref{picABagain} (a) and the theoretical Yurks-Stoler state
($|\alpha_1 >+  e^{i \frac{4}{3} \pi} |-\alpha_{1}> $).

 The generation scheme in Table \ref{YS1} does not use a coherent beam ($ |\beta_0> $).
 If we add a coherent beam ($ |\beta_0 >$), we can have precise
 control over the generated state, and  then we can increase the purity of that state.
 We have already found the
 generating condition for several Fock states  with a
 signal-to-noise ratio  higher than $1000$ \cite{youn14}.
Similarly, if we unblock the
 $b$ port in the experimental set up in  Fig.
\ref{expsetupA}, we can increase the fidelity and decrease the
absolute square of the differences of the two  Wigner functions.

 We find the generating condition for the Yurke-Stoler state with $\zeta = 0$,
 which is also known as the even-cat-state in Table \ref{YS2}.
 For the even-cat-state with $\alpha=1$, the fidelity is greater
 than $0.9999$, and the absolute square of the
difference between the  two Wigner state of the generated state and
the theoretical state ($D_w$) is less than $1.02 \times 10^{-4} $.
\begin{table}
\centering \caption{even-cat-state ($|\alpha >+  | -\alpha > $)
generating conditions.} \label{YS2}
\begin{tabular}{c|c||c|c|c|c|c|c|c|c|c|r}
\hline$ \alpha $ & $s$ & $\beta_0$  & $\gamma_0$  & $t_1$ &
$t_2$ & $\phi $ & $\theta $ &  $1-F$ & $ Log (D_w)$ & $Log(p)$  \\
\hline \hline
 1 &  0.500  & 0.341  & 0.212   & 0.832  & 0.745  & -$\pi$    & -$\pi$ & $5.1\times 10^{-5}$  &-3.99 &-2.36 \\
 \hline
 $\frac{1}{\sqrt{2}}$ &   0.455  & 0.392  & 0.181   & 0.779  & 0.593  & $\pi$   & -$\pi$& $1.2 \times 10^{-6}$    & -5.62 &-2.32 \\
 \hline
  $\sqrt{2}$ & 0. 489  & 0.208  & 0.178   & 0.930  & 0.919  & -$\pi$    & -$\pi$ & $1.2 \times 10^{-3}$   &-2.63 & -2.75  \\
  \hline
     \hline
\end{tabular}
\end{table}
\begin{figure}[htbp]
\centering
\includegraphics[width=10cm]{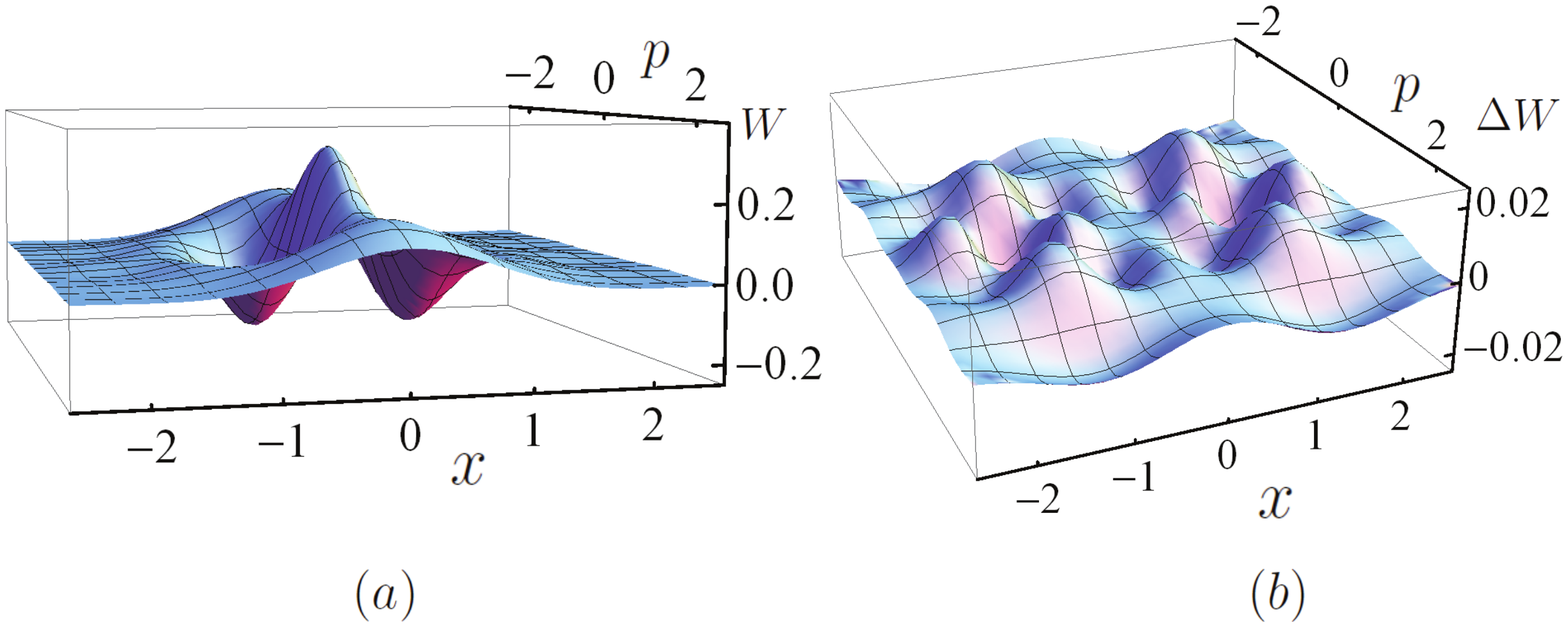} 
\caption{(a) Wigner function of the generated states ( $\alpha =
\sqrt{2} $ ) under the setups in Table \ref{YS2}. (b) Difference of
the Wigner function between the state in (a) and the even-cat-state
($|\sqrt{2} >+  |-\sqrt{2}> $) .} \label{picOneAB}
\end{figure}

 The difference between two Wigner functions is much smaller
than $ 2.4 \times 10^{-6}$ for the  even-cat-state with $\alpha=
\frac{1}{\sqrt{2}} $, and the $1-F$ becomes $1.2 \times 10^{-6} $.
Although, it is easy to generate the even-cat-state for the smaller
value of $\alpha$, the estimated purity of the even-cat-state
especially high. For the even-cat-state with $\alpha= {\sqrt{2}} $,
the estimated $D_w$ is about $2.32 \times 10^{-3} $,  and the $1-F$
becomes $1.2 \times 10^{-3} $. In Fig. \ref{picOneAB} (a), we
plotted the Wigner function of the generated state with $\alpha =
\sqrt{2} $ by the numerical parameters in Table \ref{YS2}. We
plotted the difference between two Wigner functions for the state in
Fig. \ref{picOneAB} (a) and the theoretical even-cat-state
($|\sqrt{2} > +  |-\sqrt{2}> $).

\begin{table}
\centering \caption{odd-cat-state ($|\alpha > -  | -\alpha > $)
generating conditions.} \label{YS3}
\begin{tabular}{c|c||c|c|c|c|c|c|c|c|c|r}
\hline$ \alpha $ & $s$ & $\beta_0$  & $\gamma_0$  & $t_1$ &
$t_2$ & $\phi$ & $\theta $ &   $1-F$ & $ Log (D_w)$& $Log(P)$ \\
\hline \hline
 1 &  0.500  & 0.723  & 0.012   & 0.946  & 0.921  & -$\pi$    & 0 &$1.3 \times 10^{-3}$ &- 2.58 & -2.29\\
 \hline
 $\frac{1}{\sqrt{2}}$ &   0.325  & 0.117  & 0.565   & 0.807  & 0.937  & -$\pi$    & -$\pi$  & $1.0 \times 10^{-4}$  &-3.68 &-2.44 \\
 \hline
  $\sqrt{2}$ & 0. 910  & 0.957  & 0.015   & 0.938  & 0.983  & $\pi$   & 0  & $1.0 \times 10^{-2}$  & -1.70 &-2.42 \\
  \hline
     \hline
\end{tabular}
\end{table}
\begin{figure}[htbp]
\centering
\includegraphics[width=10cm]{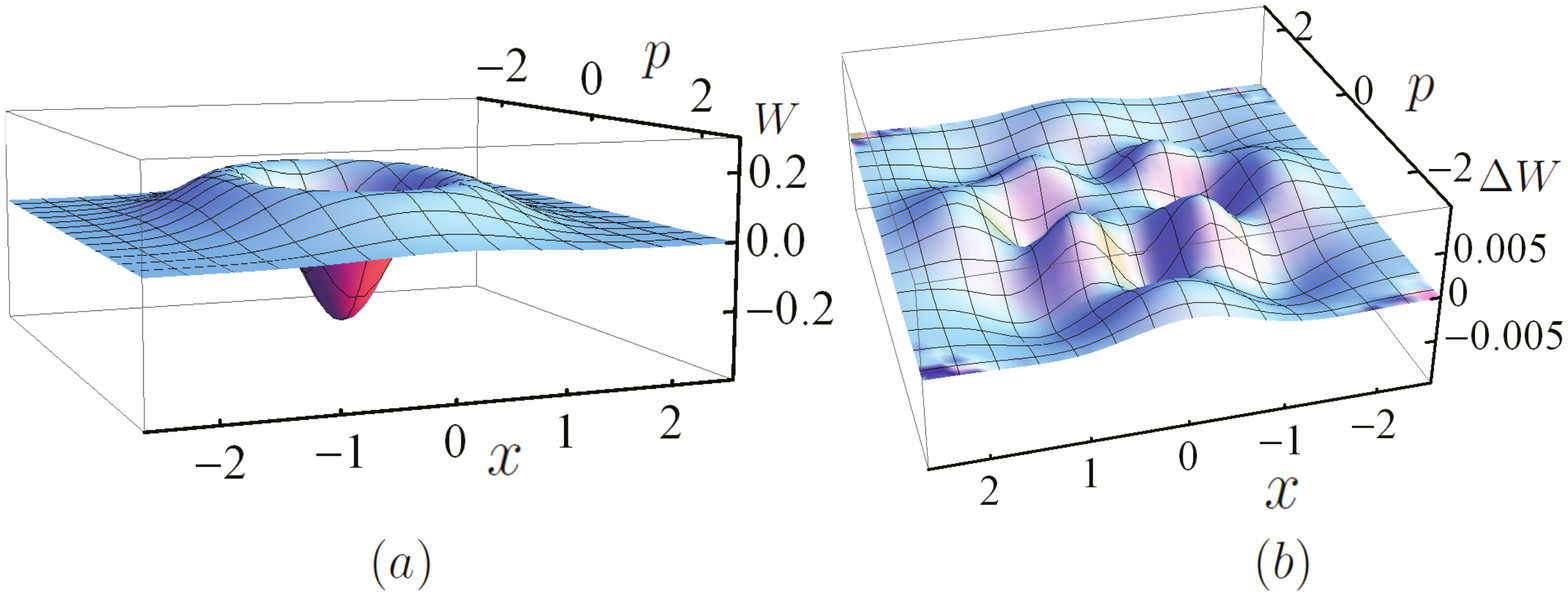} 
\caption{(a) Wigner function of the generated states ( $\alpha =
\frac{1}{\sqrt{2}} $ ) under the setups in Table \ref{YS3}. (b)
Difference of the Wigner function between the state in (a) and the
odd-cat-state ($|\frac{1}{\sqrt{2}} >- |-\frac{1}{\sqrt{2}}
> $) .} \label{picOneABodd}
\end{figure}

In our quantum state generating scheme, we also find the generating
condition for the Yurke-Stoler state with $\zeta = \pi$,
 which is also known as the odd-cat-state seen in Table \ref{YS3}.
 For the odd-cat-state with $\alpha=1$, the fidelity is greater
 than $0.9987$. Although, the generating probability  ($Log(P)$)is almost same,
  the fidelity of the generated odd-cat-state
 is relatively lower than that of the generated even-cat-state.

In Fig. \ref{picOneABodd} (a), we plotted the Wigner function of the
generated state ($\alpha = \frac{1}{\sqrt{2}} $) by using the
numerical parameters in Table \ref{YS3}, and we plotted the
difference between the two Wigner functions for the state in Fig.
\ref{picOneABodd} (a) and the theoretical odd-cat-state
($|\frac{1}{\sqrt{2}}> -  |-\frac{1}{\sqrt{2}}> $). The absolute
square of the difference between the  two Wigner states of the
generated state and the theoretical state ($D_w$) is of less than
$2.59 \times 10^{-3} $  and  $1-F$ becomes $1.3 \times 10^{-3} $.
Furthermore, the difference is much smaller than $ 2.08 \times
10^{-4}$ for the odd-cat-state with $\alpha= \frac{1}{\sqrt{2}} $
and  $1-F$ becomes $1.0 \times 10^{-4} $. For the odd-cat-state with
$\alpha= {\sqrt{2}} $, the estimated $D_w$ is of about $2.0 \times
10^{-2} $.

\section{Discussion}

 With the explicit form, the  probability
amplitude for an output state is a function of the transmittance of
two beam splitters and the amplitudes and relative phases of the
three input beams. The probabilities are calculated when the two
detectors simultaneously detect a single photon.  We have included
all of the coefficients of the input beams from zero to nine of the
number representations for three input states.

 Without a coherent state  $|\beta>$, it is possible to generate several  Yurke-Stoler
states ($|\alpha>+  e^{i \zeta} |-\alpha>$ ).   However, considering
the signal-to-noise ratio, we found that $|\beta>$ increases  the
fidelity up to $1- 5.1 \times 10^{-5} $ for the even-cat-state where
$\alpha =1 $, and the  fidelity can increase to  up to $0.999999$
for the even-cat-state with $ \alpha = \frac{1}{\sqrt{2}} $. Even
though the fidelity can decrease in the actual experiments, the
possibility to generate a high-fidelity quantum state is important
in quantum information science.  Furthermore, if we can generate a
high fidelity small cat-state, the state can be amplified through
homodyne heralding \cite{Laghaout2013}.

  Since the   probabilities have complex forms, we attempted to find
 conditions to generate  high-purity states through a numerical minimization
  method for the difference of the Wigner functions .
  In an actual experiment, $s$ was obtained as 0.63\cite{ourjou06},
   so we tried to limit the $s$ values to $1$. Considering the two coherent
  states $|\beta>$ and $|\gamma>$, the amplitudes can have large
  values in an actual experimental setup, but we tried to keep the amplitude
  to a small number.

  Our calculations are based on the assumption that we have photon-resolving
  photo detectors,  and that  we can describe our system within six photon states for each input
  state. So, we tried to keep the amplitudes $\beta_0$ and $\gamma_0$ as small
   numbers  in order to ensure the reliability of our assumptions. If the amplitudes
   are sufficiently small, we don't need  photon-resolving photo detectors.

 Considering the applicability to actual experiments \cite{Dong2014},
 if we use the input beam as a
 pulsed light with a repetition rate of
100MHz, then the generation probability of  $10^{-3}$ results in
 $10^6$ signals per second.  In actual experiments, a high
signal-to-noise ratio can be reduced as a result of  experimental
imperfections, such as mode matching and non-unity quantum
efficiency. We assumed perfect temporal and spatial mode matching
among the three input beams. These assumptions also guaranteed for
the spatial and temporal mode properties of the cat states generated
in our scheme to be well defined by the input states, and we can
precisely control the modes of the two coherent states and the
squeezed vacuum by adjusting the pump beam used to produce squeezed
states. We expect high-purity spatial and temporal modes of the cat
state.

A high-quality cat-sate can be used to study the quantum nature of
the world,
  and it is  a key
element in quantum technology.

\acknowledgements

 This study was supported by the Basic Science
Research Program, through the National Research Foundation of Korea
(NRF), funded by the Ministry of Education, Science and Technology
(20110004651)


\end{document}